\documentclass[11pt]{article}
\usepackage{geometry,amsmath}                
\usepackage{graphicx}
\usepackage{amssymb}
\usepackage{epstopdf}
\usepackage{amsthm}
\usepackage[all]{xy}
\def\tt{{\tilde{\theta}}}
\def\D{{\cal D}}
\def\R{{\cal R}}
\def\l{{l}}
\def\bR{{\mathbb R}}

\newcommand\keywords[1]{\vspace{.1in}\par\noindent{\bf Keywords}: {#1}}
\newcommand\amscode[1]{\vspace{.2in}\par\noindent{\bf 2000 AMS Subject Classification}: {#1}}
\newtheorem{example}{Example}[section]
\newtheorem{alg}{Algorithm}[section]
\title{Algebraic Methods for Inferring Biochemical Networks: a Maximum Likelihood Approach}
\author{Gheorghe Craciun\thanks{Department of Mathematics and Department of Biomolecular Chemistry, University of Wisconsin-Madison. Email: craciun@math.wisc.edu}, Casian Pantea\thanks{Department of Mathematics, University of Wisconsin-Madison. E-mail: pantea@math.wisc.edu  }, Grzegorz A. Rempala\thanks{Corresponding author. Department of Biostatistics, Medical College of Georgia, Augusta, GA 30912. E-mail: grempala@mcg.edu } }
\date{Version 1.1.1 --  10/04/08}    

\begin{document}
\maketitle
\abstract{We present  a novel  method for  identifying a  biochemical reaction network based on 
multiple sets of estimated reaction rates in the corresponding reaction rate equations  arriving from various (possibly different) experiments.  The current  method, unlike some of the  graphical approaches proposed in the literature, uses  the values of the experimental measurements only  relative to  the  geometry of the biochemical reactions  under the assumption that the underlying reaction network is the same for all the experiments. 
 The proposed approach utilizes algebraic statistical methods in order to parametrize the set of  possible reactions so as to identify the most likely  network structure, and is easily scalable to very complicated biochemical systems involving a large number of species and reactions.  The method is illustrated with a numerical example of a hypothetical network arising form a ``mass transfer"-type model.}
\keywords{Biochemical reaction network, law of mass action, algebraic statistical model, polyhedral geometry. }
\amscode{92C40, 92C45, 52B70, 62F}

\section{Introduction}

In modern biological research, it is very common to collect detailed information on time-dependent chemical concentration data for large networks of biochemical reactions (see survey papers \cite{Crampin, Maria}). 
Often, the main purpose of collecting such data is to identify the exact structure of a network of chemical reactions for which the identity of the chemical species present in the network is known  but {\it a priori} no information is available on the species interactions.  
The problem is of interest both in the setting of classical theoretical chemistry, as well as, more recently, in the context of molecular and systems biology  problems and as such has received a lot of attention in the literature  over last several decades as evidenced by multiple papers devoted to the topic \cite{Craciun_Pantea, Fay_Balogh, Himmelau_Jones_Bischoff, Hosten, Karnaukhov_etal_2007, Rudakov_1960, Rudakov_1970, Schuster_Hilgetag_Woods_Fell, Vajda_Valko_Yermakova}. 

\begin{figure}[t!] 
   \centering
   \includegraphics[width=3.5in]{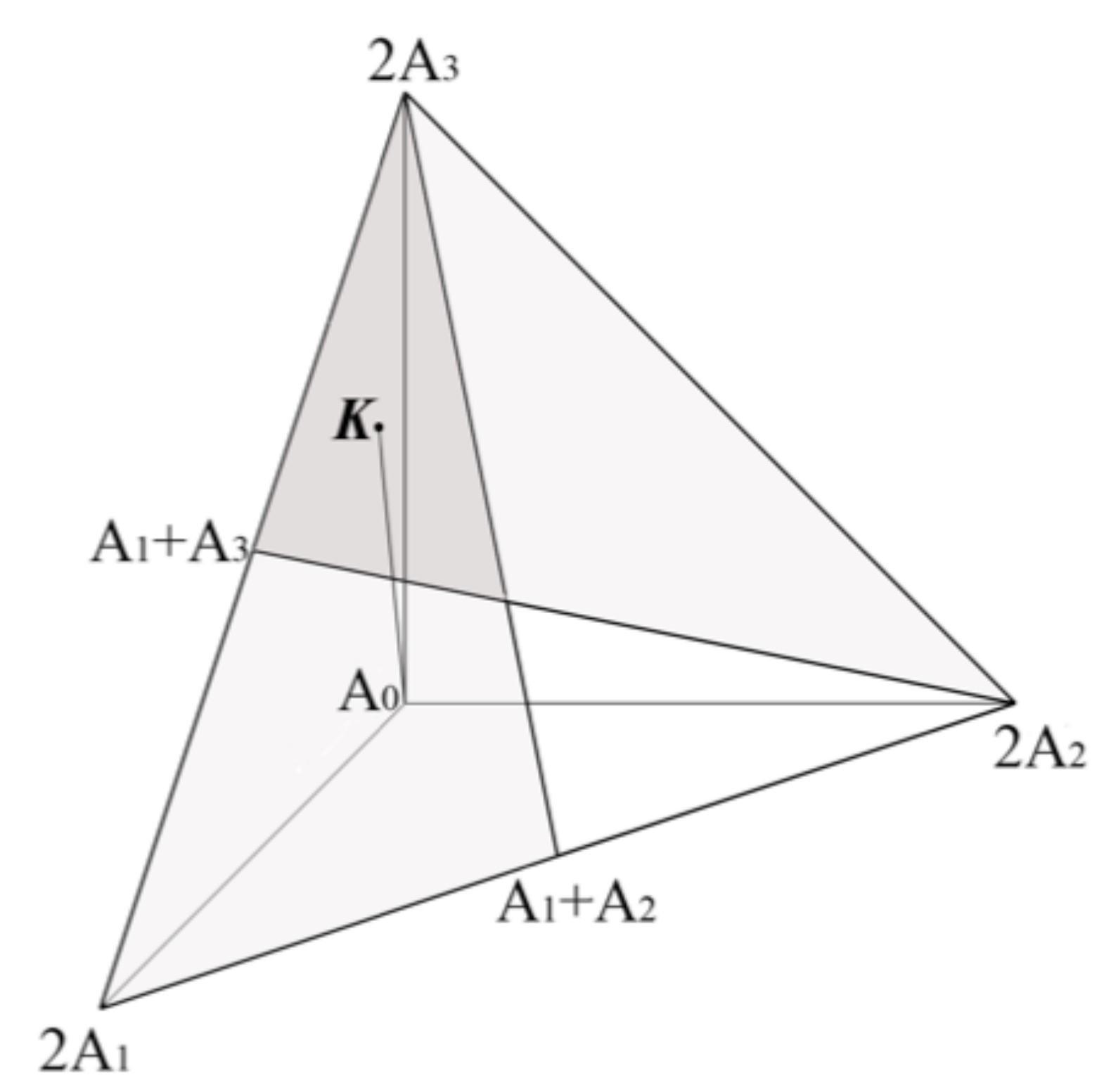} 
   \caption{\small Identifiability of reaction networks given experimental data. Note that, for a deterministic mass-action model, even if we can estimate the vector $K$ of parameter values with great accuracy, we cannot determine if the ``correct" reaction network is $\{ A_0 \to 2A_1,   A_0 \to A_1+A_2,   A_0 \to 2A_3 \}$ or  $\{ A_0 \to 2A_2,   A_0 \to A_1+A_3,   A_0 \to 2A_3 \}$, because $K$ belongs to the span of either one of these networks. However, if instead a single   point $K$ one has  available a set $\D=\{K_i, i=1\ldots,k\}$, interpreted as a result of random selection of parameter rate values according to some probability  law, then the spanning  cone of the data points may  be used  to identify the  sets of reactions that ``best explain"  the data. }
   \label{fig_1}
\end{figure}

In general, two very different reaction networks might generate identical mass-action dynamical system models, making it impossible to discriminate between them, even if one is given experimental data of perfect accuracy and unlimited temporal resolution.  Sometimes this {\it lack of uniqueness} is referred to as the ``fundamental dogma of chemical kinetics", although it is actually not a well known fact in the biochemistry or chemical engineering communities \cite{Crampin,Erdi_Toth,Epstein}. Necessary and sufficient conditions for two reaction networks to give rise to the same {\it deterministic} dynamical system model (i.e., the same {\it  reaction rate equations}) are described in \cite{Craciun_Pantea}, where the problem of identifiability of reaction networks given high accuracy  data was analyzed in detail. 
The key observation is that, if we think of reactions as vectors, it is possible for different sets of such vectors to span the same positive cones, or at least to span positive cones that have nonempty intersection (see Figure \ref{fig_1} for an example).

On the other hand, it is often the case that experimental measurements for the study of a specific reaction network or pathway are being collected under many different experimental conditions, which affect the values of reaction rate parameters. Almost always, the reactions of interest are not ``elementary reactions", for which the reaction rates parameters must be constant, but they are so called ``overall reactions" that summarize several elementary reaction steps. In that case the reaction rates parameters may reflect the concentrations of biochemical species which have not been included explicitly in the model. In such circumstances  the reaction rate parameters are {\it not} constant, but rather depend on specific experimental conditions, such as concentrations of enzymes and other intermediate species. Therefore, the estimated vector of reaction rate parameters will not be the same for all experimental conditions, but each specific experimental setting will give rise to one such vector of parameters. However, the set of all these vectors should  span a specific cone, whose extreme rays should identify exactly the set of reactions that gave rise to the data.

The purpose of the current  paper is to propose a statistical method based on the above  geometric considerations, which  allows one to take advantage of the inherent stochasticity in the data, in order to determine the {\it unique} reaction network that can best account for the results of {\it all} the available experiments pooled together. The idea is related to the notion of an {\em algebraic statistical model} (as described in \cite{Pachter_Sturmfels} Chapter 1),  and   relies on mapping the estimated reaction parameters into an
appropriate convex region of the span of reaction vectors of a network,  using  the underlying
geometry to identify the reactions which are most likely to span that region. As shown below, this approach reduces the network identification problem to a statistical  inference problem for the parameters of a multinomial distribution, which may then be solved for instance  using the classical likelihood methods.

\section{Maximum Likelihood Inference for a Biochemical Reaction Network}

In this section we develop a formal way of inferring a most likely subnetwork of a given conic network (i.e., network represented by a cone like the one in Figure~\ref{fig_1}) of the minimal spanning dimension. For the inference purpose, in the network of $m$ reactions we  assume that the empirical data $\D=\{K_i, i=1\ldots,k\}\subset  R^d$ is  available in the form of  (multiple) estimates of the  parameters of the system of differential equations corresponding to a hypothesized biochemical network.  As illustrated in Figure~\ref{fig_1}, such networks are in general ``unidentifiable" in the sense that different chemical reaction networks may give rise to the same system of differential equations. However, in the stochastic or ``statistical" sense it is possible to  identify the ``most likely'' (i.e., maximizing the appropriate likelihood  function) network  as indicated  by the   data  $\D$.

\subsection{Multinomial model }
Consider  $d$ species, and $m$ possible reactions with reaction vectors $R = \{R_1, \ldots, R_m \} \subset {\bR}^d$ among the species. (For more details about how each reaction generates a reaction vector see \cite{Craciun_Pantea}.)
Let  $\R_d$ denote the collection of all $m \choose d$ positive cones spanned by subsets of $d$ reactions in $R$.  

Denote by $cone(R)$ the positive cone generated by the reaction vectors in $R$. Let $S$ be the partition of $cone(R)$ obtained by all possible intersections of non-degenerate cones in $\R_d$.  Suppose $S$ contains $n$ full-dimensional regions $S_1, \ldots , S_n$; throughout  we shall  refer to these regions as {\it building blocks}, and to $n$ as {\it the number of building blocks}.

Let  $\Delta_{m-1}$ be a probability simplex in $\bR^m$ and let $\theta\in \Delta_{m-1}$ be a vector of probabilities associated with the reactions that give rise to $R$. We assume that these $m$ reactions have the same source complex (i.e., form a conic network), since, as explained in \cite{Craciun_Pantea}, the identifiability of a network can be addressed one source complex at a time. Define the polynomial map   \begin{align} \label{eq:gmap}
&g: \Delta_{m-1} \to \bR^n\quad \nonumber \\  \text{where}\nonumber\\&
  g_i ( \theta ) = \sum_{ C = cone(R_{\sigma(1)}, \ldots, R_{\sigma(d)}) \in {\R_d} }\frac{vol(C\cap S_i)}{vol(C)}  \theta_{\sigma(1)} \cdots \theta_{\sigma(d)}\\
  \text{for } i=1\ldots,n.\nonumber  
  \end{align}

We take\footnote{In general, it may be beneficial to consider various measures 
$vol(\cdot)$ which are absolutely continuous w.r.t.  the usual Lebesque measure. 
For instance in Section 3 we describe an example where this measure is defined via  gamma densities.}
 $\frac{vol(C \cap S_i)}{vol(C)}=0$ if $vol(C)= 0$.
 Define $s(\theta) = \sum_\sigma \theta_{\sigma(1)} \cdots \theta_{\sigma(d)}$ and 
\begin{equation}\label{eq:pmap}p(\theta)=(p_1 (\theta)\ldots,p_n(\theta))=  (g_1 ( \theta )/s( \theta ),\ldots,g_n ( \theta )/s( \theta )).\end{equation} In this setting 
  $p\in \bR^n$ is our statistical model for the data, after we substitute $\theta_m = 1 - \sum_{j=1}^{m-1} \theta_j$.  Note that we may  interpret the monomials  $\theta_{\sigma(1)}\cdots\theta_{\sigma(d)}$ in \eqref{eq:gmap} as the probabilities of a given data point being generated by the $d$-tuple of  reactions $\sigma(1),\ldots,\sigma(d)$. With this interpretation the coordinate $p_i$ of the map $p$ in \eqref{eq:pmap} is simply the conditional probability  that the data point is observed in $S_i$ given that it was generated by a  $d$-tuple of reactions.  Note that the map $p$ is  rational but, as we shall see below, the model may be re-parametrized into an equivalent one involving only the    multilinear map \eqref{eq:gmap}. 

 Let $u_i$ denote the number of data points  in $S_i$.  The log-likelihood function corresponding to a given data allocation is 
\begin{equation}\label{eq:ll}
 \l(\theta) = \sum_{i=1}^n u_i \log p_i (\theta). \end{equation}
  Our inference problem is to find 
  \begin{equation}\label{eq:max}
\hat\theta={\rm argmax}_\theta \l(\theta) \qquad \text{  subject to}\quad \sum_{i=1}^m \theta_i=1 \quad\text{and}\quad \theta_i\ge 0.
  \end{equation}
  
\begin{example}\label{ex:1}
Consider the two reaction networks described in Figure~\ref{fig_1}. 
Since the species $A_0$ does not appear as a product of any reaction, the model has effectively   $d=3$ species and  a total of $m=5$ possible reactions  $R=\{ R_1,\ldots,R_5 \}= \{ A_0 \to 2A_1,   A_0 \to A_1+A_2,   A_0 \to 2A_3,  A_0 \to 2A_2,   A_0 \to A_1+A_3    \}$.  In this case there are $n=5$ building blocks $S_1,\ldots,S_5$ defined by the intersections of all non-trivial reaction cones generated by reaction triples. Thus denoting $C_{jkl}=cone(R_j,R_k,R_l)$ for any triple $\{j,k,l\}\in\{1,\ldots,5 \}$ we have  
\begin{align*}
S_1= & C_{134}\cap C_{234}\cap C_{345}\\
S_2=& C_{134}\cap C_{145}\cap C_{234}\cap C_{245}\\
S_3=& C_{123}\cap C_{134}\cap C_{235}\cap C_{345}\\
S_4=& C_{123}\cap C_{134}\cap C_{145}\cap C_{235}\cap C_{245}\\
S_5=& C_{123}\cap C_{125}\cap C_{134}\cap C_{145}.\\
\end{align*} Note that the cones $C_{124}$ and $C_{135}$ are degenerate and are not involved in the  definitions of the $S_i$'s.  
 Denoting further $v_{jkl}^{(i)}=vol(C_{jkl}\cap S_i)/vol(C_{jkl})$ for any triple $\{j,k,l\}\in\{1,\ldots,5 \}$,  we see that the the map  \eqref{eq:gmap} becomes   
\begin{align*}
g_1 ( \theta ) =& v_{134}^{(1)}  \theta_1\theta_3\theta_4 +v_{234}^{(1)}\theta_2\theta_3\theta_4 +v_{345}^{(1)}  \theta_3\theta_4\theta_5 \\
g_2 ( \theta ) =& v_{134}^{(2)}  \theta_1\theta_3\theta_4 +v_{145}^{(2)}\theta_1\theta_4\theta_5 +v_{234}^{(2)}  \theta_2\theta_3\theta_4+v_{245}^{(2)}  \theta_2\theta_4\theta_5 \\
g_3 (\theta) =& v_{123}^{(3)}  \theta_1\theta_2\theta_3 +v_{134}^{(3)}\theta_1\theta_3\theta_4 +v_{235}^{(3)}  \theta_2\theta_3\theta_5 \\
g_4 (\theta) =& v_{123}^{(4)}  \theta_1\theta_2\theta_3 +v_{134}^{(4)}\theta_1\theta_3\theta_4 +v_{145}^{(4)}  \theta_1\theta_4\theta_5+v_{235}^{(4)}\theta_2\theta_3\theta_5 +v_{245}^{(4)}  \theta_2\theta_4\theta_5 \\
g_5 (\theta) =& v_{123}^{(5)}  \theta_1\theta_2\theta_3+\theta_1\theta_2\theta_5 +v_{134}^{(5)}\theta_1\theta_3\theta_4 +v_{145}^{(5)}  \theta_1\theta_4\theta_5, \\
\end{align*}
where  the coefficients satisfy   $\sum_i v_{jkl}^{(i)}=1$ for any triple $\{j,k,l\}$ appearing on the right-hand-side  in the formulas above. 
 The rational map \eqref{eq:pmap} is therefore given by $$p=\frac{g}{\sum_{jkl} \theta_j\theta_k\theta_l}$$ where the sum in the denominator extends over all distinct triples $\{j,k,l\}$ excluding  $\{1,2,4\}$  and $\{1,3,5\}$, i.e., the ones corresponding to degenerate cones. 
\end{example}

\subsection{Multilinear representation}
The model representation via a rational map \eqref{eq:pmap} may be equivalently described in terms of a  simpler  polynomial  map  \eqref{eq:gmap} as follows. 
Let  us substitute   $\tt_i=\theta_i s^{-1/d}$ for $i=1,\ldots m$ and define 
$$\tilde{g}_i(\tt)=p_i(\theta).$$
Note that $\tilde{g}_i:{\bR}^m_{>0}\to {\bR}^n$ and  $l(\tt)=l(\theta).$
Thus we may consider  a following more convenient version of \eqref{eq:max}. Find 
\begin{align*}\label{eq:max2}
&\hat\theta={\rm argmax}_{\tt}\ \l(\tt)\\
&\text{subject to}\quad  \sum_\sigma \tt_{\sigma(1)} \cdots \tt_{\sigma(d)}= \sum_i \tilde{g}_i(\tt)=\sum_{i=1}^n p_i(\theta)=1, \quad \forall_i\ \tt_i\ge 0.\tag{$\ast$}
\end{align*}

Consider  a fixed $d$-tuple  of reactions  (say, $\sigma_1$) and in the formulas 
for $\tilde{g}_i$ ($i=1\ldots,n$)   substitute     $\tt_{\sigma_1(1)} \cdots \tt_{\sigma_1(d)}=1-\sum_{\sigma\ne \sigma_1} \tt_{\sigma(1)} \cdots \tt_{\sigma(d)}$.  Note that the  resulting  algebraic statistical map is multilinear i.e, linear in one parameter $\tt_k$ when all others are    fixed. 
For instance,  as a function of $\tt_1$ we have 
$$p_i(\tt_1|\cdot)=  a_{i}\tt_1+b_i \quad i=1\ldots,n $$
where $\sum_i a_i=0$ and $\sum_i b_i=1$ and $a_i, b_i$ are given in terms  of $\tt_l$ for $l>2$. 
  
By Varchenko's theorem (see, \cite{Pachter_Sturmfels} chapter 1) the conditional, one dimensional version of problem ($\ast$) may be now solved iteratively for  each $p_i(\tt_1|\cdot),$ $i=1,\ldots,n$ by finding a unique root of the score equations in the regions bounded by the ratios $-b_i/a_i$. 

\bigskip

\noindent \textbf{Maximization algorithm.} Due to  the conditional convexity of the one dimensional problems 
the above considerations suggest that the following algorithm for (local) maximization of $\l(\tt)$ should be valid  (cf. also \cite{Pachter_Sturmfels}, Example 1.7, page 11): 
\begin{alg}\label{alg:1} \ \\ \vspace{-.2in}
\begin{enumerate}
\item Pick  initial vectors $\tt$ and  $\tt_{old}\in {\bR}^m.$
\item While  $|l(\tt)-l(\tt_{old})|>\epsilon$ 
\begin{itemize}
\item  $\tt_{old}\leftarrow \tt$  
\item for k=1 to m do 
\begin{itemize}

\item compute $a_i, b_i$ (as functions of $\tt_j ,j\ne k$)  
\item identify the   bounded interval as determined by  Varchenko's fromula  which is statistically meaningful (there is only one).
\item use a simple hill-climbing algorithm  to find an optimal   $\tt_k^{opt}$ in that interval
\item update $\tt_k\leftarrow\tt_k^{opt}$
\end{itemize} 

\end{itemize}
\item Recover $\theta$ from $\tt$ by taking $\theta_k=\tt_k /\sum_i \tt_i$. 
\end{enumerate}
\end{alg}

The advantage of the algorithm  above is that  it reduces a potentially very complicated   multivariate optimization problem in which  $d$  and $m$ are large to  iteratively  solving of  a simple,   univariate one.  The disadvantage is that due to its dimension-iterative character the algorithm is seen to be slow and for smaller networks perhaps  less efficient than some  off-the-shelf optimization algorithms available in commercial software (e.g., some modified hill-climbing methods with random restarts). For that reason in our numerical example below we used the standard Matlab optimization package rather than Alg.~\ref{alg:1}.
  
 In the reminder of the paper we revert to the notation of Section~1 and the original problem \eqref{eq:max}.  Based on $(\ast)$ in this section   we may thus extend map $g$ to $\bR^m_{>0}$, take $s(\theta)=1$   in \eqref{eq:ll}  and re-cast the original likelihood maximization problem  \eqref{eq:max} as 
  \begin{equation*}\label{eq:max3}
\hat\theta={\rm argmax}_\theta \sum_i u_i \log g_i (\theta) \qquad \text{subject to}\quad \sum_\sigma \theta_{\sigma(1)}\cdots \theta_{\sigma(d)}=1 \quad\text{and}\quad \theta_i\ge 0\eqno{(\ref{eq:max}^\prime)}
  \end{equation*} where the $g_i$'s are given by \eqref{eq:gmap}.

\section{Simulated Numerical Example}

In this section we illustrate the ideas discussed above by analyzing a specific numerical example in detail. 

If we have $d$ chemical species and data of the form $\D=\{K_i, i=1\ldots,k\}$, then we would hope that the statistical algorithm described above should recover the most likely $d$ reactions out of a given list of $m\geq d$ possible reactions, by finding the maximizing vector $\hat{\theta}$ of  the corresponding log-likelihood function. In what follows the setup of the problem is that of (\ref{eq:max}$^\prime$). To this end,  consider   the following four-dimensional example: 
\begin{equation}\label{CRN1}
\xy ;<1pc, 0pc>:
\POS(0,0)*+{A_{0}}
\ar@<.3 ex> @{->}        +(-3,-3)*+{A_{1}+A_2}
\ar@<.3 ex> @{->}        +(-3,3)*+{A_{1}}
\ar@<.3 ex> @{->}        +(3,3)*+{2A_3}
\ar@<.3 ex> @{->}        +(3,-3)*+{A_2+A_3}
\endxy
\end{equation}
where $A_{i}$, $i = 0,1,2,3$, denote four chemical species. We shall use the above  reaction network to simulate ``experimentally measured" data and to test  the performance of our method outlined in Section~2. To this end  we shall augment the above  network by including one or more ``incorrect'' reactions, and shall  check whether our likelihood-based algorithm (\ref{eq:max}$^\prime$) is  able to identify the original ``correct" set  of four reactions.

\bigskip
 
\begin{figure}[b!] 
   \centering
   \includegraphics[width=4.5in]{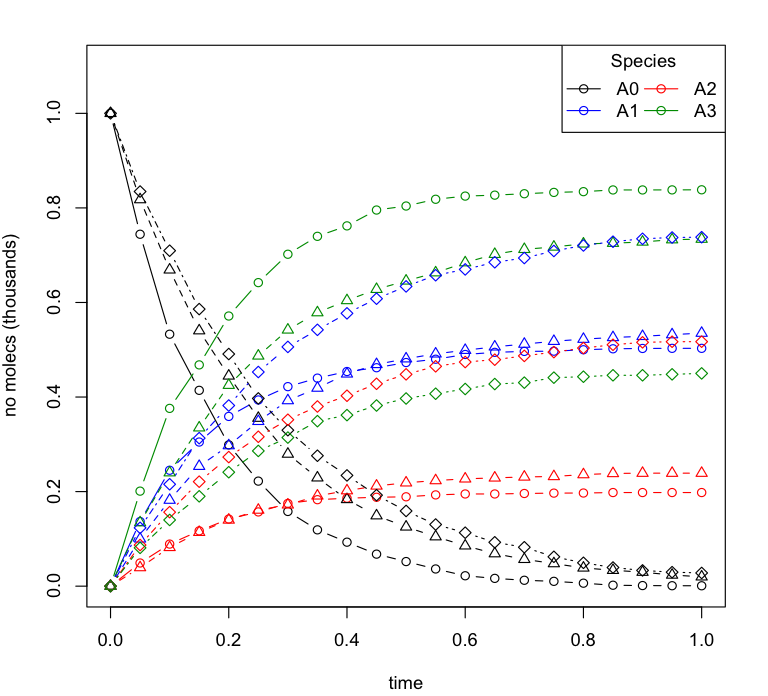} 
   \caption{\small An example of generation of the data points $K_i \in\D$  for $i=1,2,3$ via a two-step process of simulation and estimation. Three stochastic trajectories  of the reaction network  \eqref{CRN1} were simulated via Gillespie algorithm with propensity (reaction) constants drawn randomly according to gamma~$G(1.5,1)$ distribution. The trajectories values at the  data collection points are marked at 20 equally-spaced time-points from 0 to 1.  The data from the set of trajectories was used in order to estimate the coordinates for each  set of coordinates of $K_i$, $i=1,2,3.$  The numerical values of the reaction rates corresponding to the given trajectories along with their least-squares estimates are presented in Table~\ref{tab:1}.}
   \label{fig:trajs}
\end{figure}

\noindent \textbf{Data generation.} Note that the (deterministic)  dynamics of the chemical reaction network (\ref{CRN1}) is governed by the linear differential equations of the form \begin{equation}\label{eq:rre} dA_i/dt=\gamma_i A_0\qquad i=0,\ldots,3.\end{equation} Thus in our example each data point $K_i\in \D$ $(i=1,\ldots,k)$ was generated by estimating the set of parameters $(\gamma_i)$ of  the true reaction network (\ref{CRN1}). For each one of the $k$ data points the set of parameters $(\gamma_i)$ was drawn independently  from a gamma distribution $G(\alpha,\lambda)$ with parameters $\alpha=1.5$ and $\lambda=1$.  In order to  identify the coordinates of the  points in  $\D$, the estimated parameters $\hat{\gamma}_i$, $i=0,1,2,3,$ were calculated each time by fitting the trajectories \eqref{eq:rre} to the time series data points generated from the stochastic process  tracing \eqref{eq:rre} (see \cite{EK86}). The  Gillespie algorithm (see \cite{RRK}) was used to generate the 20 equally-spaced values of the trajectory of random process on the interval $(0,1)$ with the fixed initial condition. An example of  three random trajectories with independently generated reaction constants values is given in Figure~\ref{fig:trajs}. These three trajectories would give rise to three  independently estimated sets of values $(\gamma_i)$ and consequently to three data points $K_i\in \D$. 
The fitting was based  on  the least-squares criterion which is statistically  justified for estimation purpose of $(\gamma_i)$ in this particular case by an appropriate central limit theorem (cf. e.g., \cite{EK86} chapter 11). In our example, rather than the conditional Algorithm~\ref{alg:1},  we have implemented a more widely used  local optimization procedure with random restarts  as offered by  the Matlab function \texttt{lsqcurvefit}. 

A resulting single data point $K_i=\hat{\gamma}_i$ is the statistical least-squares estimate of a realization of four  independent gamma-variates  $G(1.5,1)$  which may  be viewed as coordinates of the true reaction constants  vector   in the species coordinate system $[A_0, A_1, A_2, A_3]$. This representation of data points is not related to a choice of the reactions; note, however, that each data point\footnote{Here we assume tacitly that the estimation error is sufficiently small and that the statistical estimation procedure is consistent. It turns out this is typically  the case in the settings similar  to our simulated example, but the discussion of  the precise conditions under which this is true in real experimental settings goes beyond the scope of our present discussion. For our current example  a brief inspection of the Table~\ref{tab:1} indicates a reasonably good agreement between the estimates and the true values of the reaction rates both in terms of actual values as well as the corresponding  SE's.} lies inside the open convex cone generated by the true reactions  (\ref{CRN1}). As shown in \cite{Craciun_Pantea} the coordinates of  $K_i$ in the basis given by the reaction vectors in  (\ref{CRN1}) are precisely the estimates of the true rate constants.

\begin{table}[t!]\label{tab:1}
\begin{center}
\begin{tabular}{l|c|c|c}

Reaction &  Propensity Constants & Estimated Values &  Estimators SEs\\
\hline
$A_0\rightarrow A_2+A_3$&(0.953, 0.630, 0.065) & (0.898, 0.668, 0.056) &(0.091, 0.066, 0.032)\\
$A_0\rightarrow A_1$& (2.982, 1.869, 0.711)&(2.711, 1.905, 0.876)&(0.136, 0.098, 0.060)\\
$A_0\rightarrow A_1+A_2$& (0.328, 0.336, 1.740)&(0.301, 0.349, 1.874)&(0.058, 0.048, 0.087)\\
$A_0\rightarrow 2A_3$&(1.996, 1.262, 0.853)&(2.064, 1.212, 0.824)&(0.118, 0.075, 0.058)\\

\end{tabular}\caption{\small Three sets of reaction kinetic constants corresponding to the trajectories depicted in Figure~\ref{fig:trajs} along with their estimated values (obtained via least-squares fitting) and standard errors of the estimates. } 
\end{center}
\end{table}
The  data set $\D=\{K_i, i=1\ldots,k\}$ used in the simulation described above was based on  considered $k=50$ data points. The  first three data points are summarized in Table~\ref{tab:1}. 

In order to test our method, let us first add one incorrect reaction, $A_0\to A_2$, and from this point on suppose we have no {\it a priori} knowledge of the true chemistry; therefore, the five possible reactions are as follows:  

\begin{equation}\label{CRN2}
\xy ;<1pc, 0pc>:
\POS(0,0)*+{A_{0}}
\ar@<.3 ex> @{->}        +(-3,-3)*+{A_{1}+A_2}
\ar@<.3 ex> @{->}        +(-3,3)*+{A_{1}}
\ar@<.3 ex> @{->}        +(0,3)*+{2A_3}
\ar@<.3 ex> @{->}        +(3,3)*+{A_2}
\ar@<.3 ex> @{->}        +(3,-3)*+{A_2+A_3}
\endxy
\end{equation}

Later in this section we also consider the case where we add not just one, but several incorrect reactions.

\bigskip 

\noindent \textbf{Calculation  of the log-likelihood function.}   In order to obtain an estimate $\hat{\theta}$  via $(\ref{eq:max}^\prime$) one needs to be able to evaluate the map (\ref{eq:gmap}), i.e., in addition to the data counts  vector $u\in {\bR}^n$ in  (\ref{eq:ll})  one also  needs  to know the values of the coefficients of the polynomial map. Whereas the calculation of the exact values  is difficult for $d>2$, one may typically resort to Monte-Carlo approximations (see, e.g.,  \cite{HY08}).  In our current example, for a non-degenerate (i.e., 4-dimensional) cone  $C$, we have computed the approximate relative volumes $vol(C\cap S_i)/{vol(C)}$ using the following Monte Carlo method. For each cone $C$ we generated $N=2000$ points inside $C$ with the corresponding conical coordinates randomly drawn from  the four independent gamma$G(1.5,1)$ random variable and then  counted the proportion of the total points  falling into $C\cap S_i$ i.e., used the approximation 
$$ \frac{vol(C\cap S_i)}{vol(C)}\approx(\#\text{points in }C\cap S_i)/N\quad  i=1\ldots, n.$$ 
With the coefficient values determined as above, the coordinate polynomial maps ${g}_i$ in \eqref{eq:gmap} were easily calculated now by identifying the cones that contained the appropriate building block regions $S_i.$

\begin{figure}[h]
\begin{minipage}[t]{7cm}
\begin{center}
\includegraphics[width=7cm,clip]{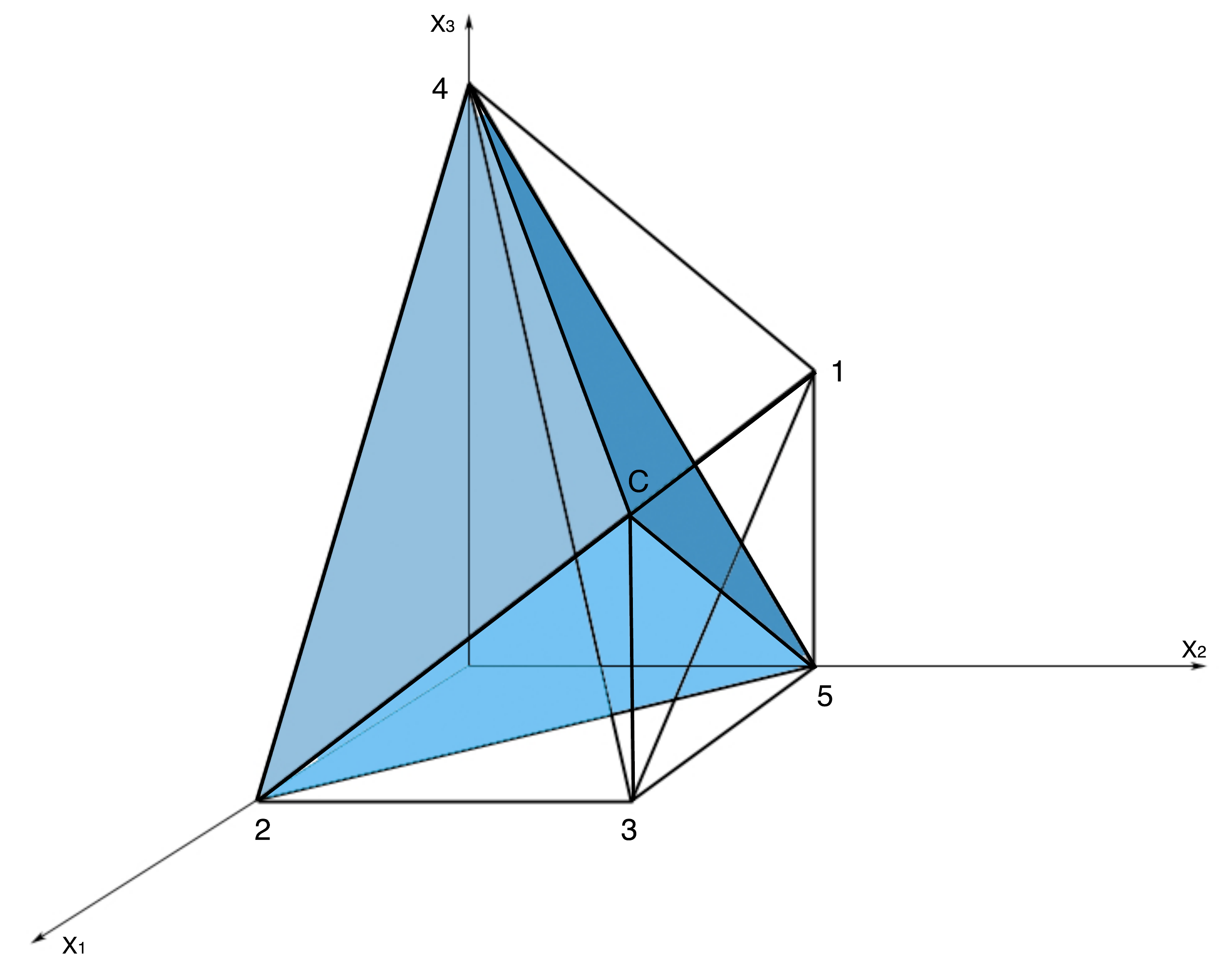}
\caption[Short]{\label{Fig1} \small Geometry of building blocks for reaction network (\ref{CRN2}).}
\end{center}
\end{minipage}
\hskip0.5cm
\begin{minipage}[t]{7cm}
\begin{center}
\includegraphics[width=7cm,clip]{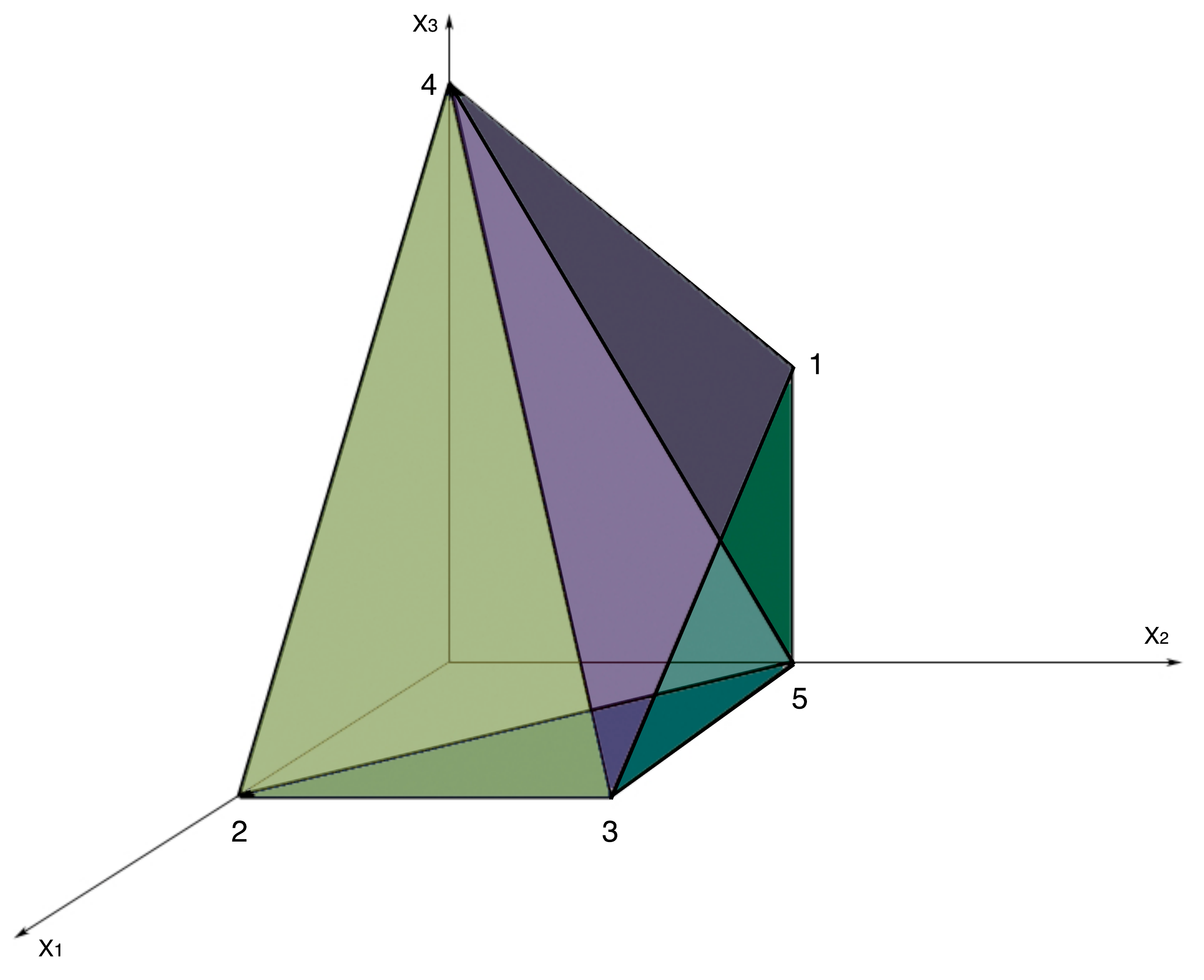}
\caption[Short caption for figure 2]{\label{Fig2} \small Faces of polyhedron corresponding to reaction network (\ref{CRN2}).}
\end{center}
\end{minipage}
\end{figure}

\vspace{.2cm}

\bigskip

\noindent{\bf Visualization of  the chemical  network.} The geometry in (\ref{CRN2}) can be visualized in the 3-dimensional subspace $\mathcal{W} \subset {\bR}^n$ generated by $\{A_1, A_2, A_3\}.$ This follows  as all the reaction targets are in this subspace, and we can understand the configuration of relevant four-dimensional  cones by looking at their intersections with $\mathcal{W}$. Each four-dimensional cone with vertex $X_0$ intersects $\mathcal{W}$ along a tetrahedron. The intersections of all these tetrahedra cut out the building blocks corresponding to our example (\ref{CRN2}), as illustrated in Figures \ref{Fig1} and \ref{Fig2}. There are five vertices labeled by numbers corresponding to the five target reactions in (\ref{CRN2}); they form a six-faced convex polyhedron $\mathcal{P}$. Let $C$ be the intersection of line passing through points 1 and 2, denoted $(12)$, with the plane $(345)$. Then all building blocks are tetrahedra with a vertex at $C$ and the opposite face being one of the six faces of the polyhedron $\mathcal{P}$. For example, the building block $(C245)$ is depicted in Figure~\ref{Fig1}.

Not surprisingly,  the 50 data points generated in our example were  distributed among the building blocks that compose the tetrahedron $(1234)$ corresponding to the true reactions; 32 data points fell inside the building block $(C234)$ and 18 inside $(C134).$ The log-likelihood function was found  in this case as 
$$\l(\theta) = 32\log(.706\cdot\theta_1\theta_2\theta_3\theta_4+.35\cdot\theta_2\theta_3\theta_4\theta_5)+18\log(.294\cdot\theta_1\theta_2\theta_3\theta_4+.339\cdot\theta_1\theta_3\theta_4\theta_5).$$

\bigskip

\noindent\textbf{Maximization of log-likelihood.} In order to maximize $l(\theta)$ or, equivalently, to minimize $-l(\theta)$, we used the Matlab function \texttt{fmincon} for constrained optimization. As in (\ref{eq:max}'), the constrain is given by the condition $s(\theta) = \sum_{\sigma} \theta_{\sigma(1)}\cdots\theta_{\sigma(4)} = 1$ and comes from the fact that there are 4 reactions in the true network. This constrain also assures that $\tilde g$ maps into the probability simplex $\Delta_{n-1}$, i.e., defines an algebraic  variety (polynomial map) which corresponds to a valid tstatistical model. 

The optimization is repeated $2^{m-1}$ times (i.e., 16 times for the example for network (\ref{CRN2})) with random initial conditions satisfying the constrain. A list of (local) minima  was created and entries were merged if they were very close. The point $\theta$ that achieved the smallest local minimum was reported together with the percentage of time the algorithm ended up at that particular point (success rate). The output for example (\ref{CRN2}) given by the customized  Matlab function was 

\vspace{.2cm}
\noindent \texttt{Minimum of negative log-likelihood: 33.19\\  Theta:\\
1\hspace{2cm}1\hspace{2cm}1\hspace{2cm}1\hspace{2cm}0.\\
Hits: 16 out of 16, 100\%.}
\vspace{.2cm}

\noindent  As we may  see from these results, in  the notation of Figures \ref{Fig1} and \ref{Fig2} the algorithm identified the true reactions (targets) 1, 2, 3 and 4 and discarded the incorrect reaction~5.

\bigskip
 
\noindent\textbf{More numerical comparisons.} We also ran example (\ref{CRN1}) with, respectively,  two, three, four and five incorrect reactions added  to the set of four correct ones. The true network was always identified and the success rate (percentage of correct hits for various random initial guess) was high. The  results of these additional experiments are summarized in Table~2.

\begin{table}[b!]
\begin{center}
\begin{tabular}{c|c|c|c|c|c}
\# reactions & \# cones& \# non-degenerate & \# building & running&success\\
	              &               & cones                       &  blocks                              & time          &rate        \\
\hline
m=5               &5            &5				    &6				    &7 sec		 &100\%\\
m=6               &15          &11		             &15		             &29 sec		 &94\%\\             
m=7               &35          &30			    &133			   &6 min 32 sec &94\%\\
m=8               &70          &64			    &871		             &1h 12 min	 &98\%\\
m=9               &126        &115			    &2397		             &8h 55 min	 &96\%\\
\end{tabular} 
\end{center}
\caption{\small Summary of numerical results for $k=50$ data points, using $N=2000$ rays in the Monte Carlo relative volume computation, $2^{m-1}$ optimizations with random initial guess. The code was tested on a 2.8 Ghz Intel Core2Duo iMac machine.}
\end{table}

\section{Summary and Discussion}

We have proposed herein a statistical method for inferring a biochemical reaction network  given several sets of data that originate from ``noisy" versions of the reaction rate equations associated with the network. As illustrated in some earlier work of some of the authors \cite{Craciun_Pantea}, in the usual deterministic sense such networks are in general unidentifiable, i.e., different chemical reaction networks may give rise to exactly the same reaction rate equations.  In practice, the matters are further complicated since the coefficients of the reaction rate equations are estimated from available experimental data, and hence are subject to measurement error and, moreover,  their actual values may differ at different experimental conditions, i.e. at different data points.  
The statistical approach described here is largely unaffected by these problems, as it only relies on the geometry of the network relative to the data distribution, in order to identify the  sets of most likely reactions. Hence, the method takes advantage of  the algebraic and geometric representation of the network rather than merely the observed experimental values of the network species, as is  commonly the case in  network inference models based on  graphical methods, like e.g. Bayesian or probabilistic boolean networks. Still, in order to use the proposed multinomial parametrization of a biochemical network, the method does require a valid way of mapping the experimentally estimated rate coefficients into the networks' appropriate convex regions, and with very large measurement errors is likely to perform poorly. On the other hand, precisely because of the need for the experimental data  mapping, the method has a very attractive feature of being able to potentially combine   variety of different data sets obtained by various methods into one set of experimental points placed in a convex hull of the network building blocks. These universality properties of the method  require further studies and possibly a development of additional statistical methodology beyond the scope of our present work. In the current paper our main goal was to present a proof-of-concept example based on simulated data, with a purposefully straightforward but non-trivial model discrimination problem. For the example provided in this paper the method was seen to perform very well, with almost perfect discrimination against incorrect models even as the complexity of the model selection problem increased.

Nonetheless, further studies and developments are needed to assess how well the method may perform on more challenging and realistic data sets. In particular, one of the aspects of the methodology which was not pursued here, and which could  improve its computational scalability, is the utilization of techniques from computational algebra in order to increase the efficiency and further automate the proposed maximization algorithm. 

\bigskip

\noindent {\bf Acknowledgements.} The authors would like to thank Peter Huggins and Ruriko Yoshida for very helpful discussions, and additionally thank Peter Huggins for making available his Matlab script for volume calculations. The research was partially sponsored by the  ``Focused Research Group" grants NSF--DMS 0840695 (Rempala) and  NSF--DMS 0553687 (Craciun) as well as by the NIH grant 1R01DE019243-01 (Rempala).

\end{document}